\documentclass[letterpaper, 10 pt, conference]{ieeeconf}  

\IEEEoverridecommandlockouts                              

\usepackage{graphics} 
\let\labelindent\relax
\usepackage{framed,enumitem} 
\usepackage{epsfig} 
\usepackage{mathptmx} 
\usepackage{times} 
\usepackage{amsmath} 
\usepackage{amssymb}  
\usepackage{arydshln}

\usepackage[noadjust]{cite}
\title{\LARGE \bf
LPV sequential loop closing for high-precision motion systems*
}

\author{Yorick Broens, Hans Butler and Roland T\' oth
\thanks{*This work has received funding from the ECSEL Joint Undertaking (JU) under grant agreement No 875999 and from the Ministry of Innovation and Technology NRDI Office within the framework of the Autonomous Systems National Laboratory Program.}
\thanks{Y.Broens, H.Butler and R.T\'oth are with the Department of Electrical Engineering, Eindhoven University of Technology, Eindhoven, The Netherlands. H.Butler is also affiliated with ASML, Veldhoven, The Netherlands. R.T\'oth is also affiliated with the Systems and Control Laboratory, Institute for Computer Science and Control, Hungary,  ({\tt\small email: Y.L.C.Broens@tue.nl}). }
}

\begin{document}

\maketitle
\thispagestyle{empty}
\pagestyle{empty}

\begin{abstract}
Increasingly stringent throughput requirements in the industry necessitate the need for lightweight design of high-precision motion systems to allow for high accelerations, while still achieving accurate positioning of the moving-body. The presence of position dependent dynamics in such motion systems severely limits achievable position tracking performance using conventional sequential loop closing (SLC) control design strategies. This paper presents a novel
extension of the conventional SLC design framework towards linear-parameter-varying systems, which allows to circumvent limitations that are introduced by position dependent effects in high-precision motion systems. Advantages of the proposed control design approach are demonstrated in simulation using a high-fidelity model of a moving-magnet planar actuator system, which exhibits position dependency in  both actuation and sensing.
\end{abstract}

\section{Introduction}
\label{Section:Introduction}
Growing demands in the industry result in increasingly stringent requirements on throughput and positioning accuracy of motion systems, such as wafersteppers, scanners, pick-and-place machines and wire bonders, see \cite{Butler,411117,HEERTJES20161, 6225187}. Traditionally, control design for multiple-input multiple-output (MIMO) systems is simplified by excellent mechanical design which ensures high stiffness and reproducibility of the design, see \cite{Oomen}. Moreover, highly stiff mechanical designs result in motion behavior dominated by rigid body dynamics which simplify the control design procedure and allow to use methods like rigid body decoupling to efficiently handle MIMO systems, see \cite{Steinbuch2013}.  In many industrial applications, \emph{sequential loop-closing} (SLC) strategies are applied for rigid body feedback control design due to several (practical) advantages. First,
these methods employ well-understood control design strategies, such as loop-shaping techniques (see \cite{190Steinbuch}, \cite{HOVD19941601}). Additionally, the SLC framework allows for feedback control design using a \emph{non-parameteric model} of the motion system, e.g. frequency response function (FRF) measurements, thus circumventing the necessity of identifying an accurate dynamic model of the system that can capture complicated high-frequent resonance dynamics.

However, the ever increasing throughput demands in the industry necessitate the design of lightweight mechanical structures to allow for ultra-high accelerations, while keeping the power demands relatively low. Therefore, increased throughput is obtained at the cost of introducing low-frequent flexible dynamics due to the limited stiffness of the mechanical structure, see \cite{VibrationModes}.

Additionally, control design of high-precision motion systems is further complicated by the presence of position dependent effects. For many high-precision motion systems, position dependent resonance dynamics are introduced by relative displacements of the moving-body with respect to the sensor frame and/or actuation frame, resulting in the need for coordinate frame transformations to relate the input forces and/or measurement signals to the center of mass of the moving-body.

In general, the presence of flexible dynamics severely limits the achievable feedback control bandwidth (see \cite{STEINBUCH1998278}), which is even more critical in case they are position dependent, since position dependent effects manifest as dynamic uncertainties. While previously SLC control design strategies handled position dependency in terms of robustness of the resulting LTI controller, the performance price of robustness in case of the novel lightweight designs would be intolerable in practise. 

In order to circumvent the aforementioned performance limitations, a novel control design approach is presented, which extends the conventional \emph{linear-time-invariant} SLC method (e.g. see \cite{skogestad2007multivariable}) to \emph{linear-parameter varying} (LPV) systems. The proposed LPV SLC control design strategy preserves the advantages of the traditional SLC approach, while allowing for increased closed-loop performance for motion systems affected by position dependency. The presented control design approach is based on local controller design with loop-shaping strategies, thus allowing for an interpolation based parameterization of the controller coefficients on the scheduling vector respectively. Additionally, an efficient implementation strategy is presented, which allows for real-time implementation of the designed position dependent feedback controllers.

The main contributions of this paper are:
\begin{itemize}
  \item[(C1)] The development of an extension of the SLC control design approach, such that position dependency is explicitly taken into consideration, thereby allowing for additional degrees of freedom to the control design, such that increased closed-loop performance is achieved.
  \item[(C2)] The development of an efficient implementation strategy for the position dependent controllers, which allows for real-time implementation of the proposed control concept.
\end{itemize}

This manuscript is organized as follows. First, the problem formulation is presented in Section \ref{Section_ProblemFormulation}. Next, Section \ref{Section_LPVSLCframework} presents the conventional sequential loop-closing framework for rigid-body feedback control design for high-precision motion systems. In Section \ref{LPVExten}, the proposed LPV SLC framework approach is introduced together with an efficient implementation strategy for real-time application of the resulting position dependent controllers. Section \ref{Section_Academicexample} provides a simulation study of the proposed LPV SLC control approach on a high-fidelity model of a state-of-the-art moving-magnet planar actuator.
Finally, in Section \ref{Section_Conclusion}, conclusions on the proposed methodology are drawn.
\section{Problem formulation}
\label{Section_ProblemFormulation}
\subsection{Background}
\begin{figure}[t]
    \centering
    \includegraphics[trim={2.9cm 0cm 3.4cm 1.4cm},clip,width=\linewidth,height = 5.5cm]{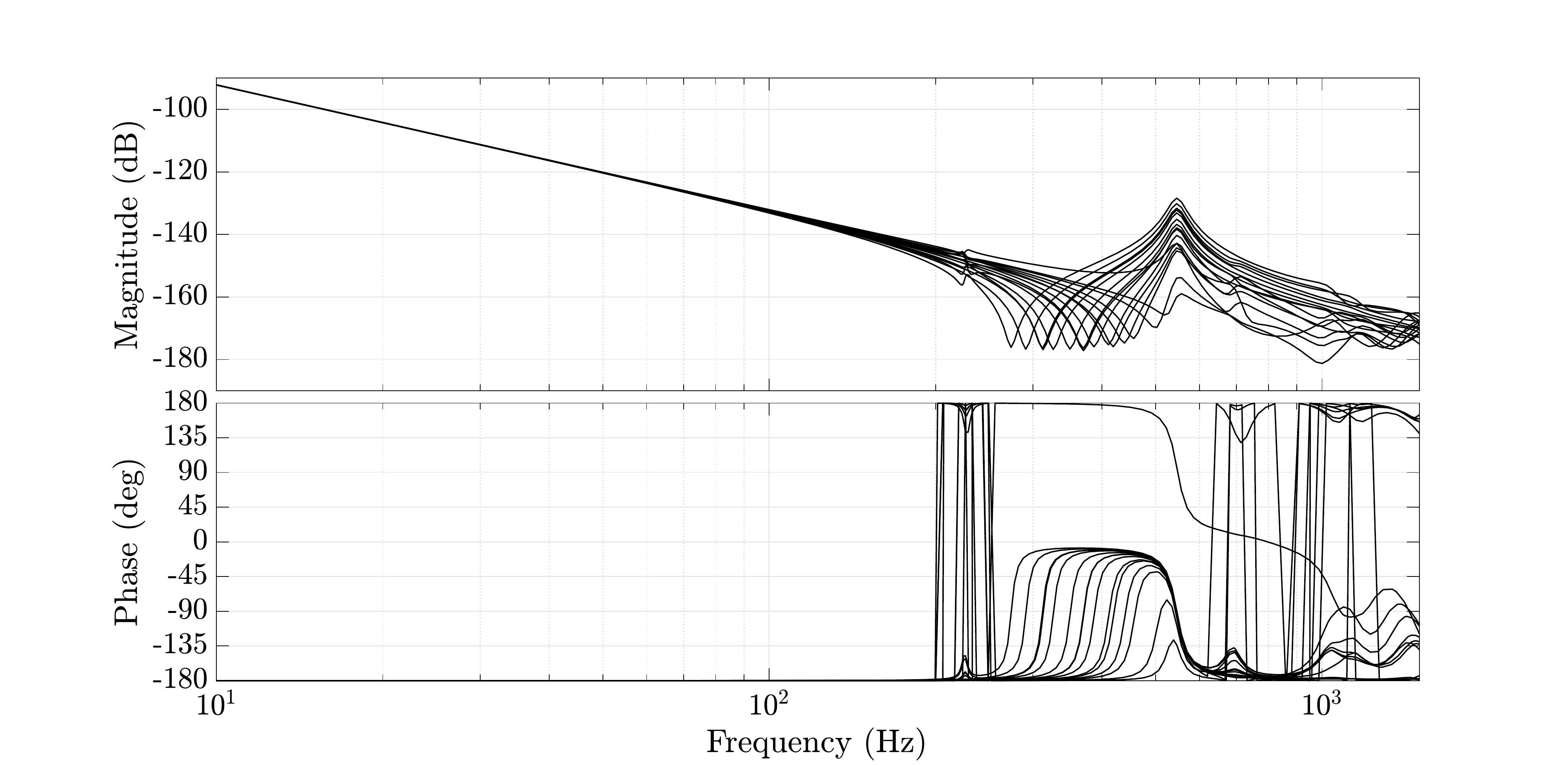}
    \caption{Low-frequent position dependent resonance dynamics of a high-precision moving-magnet planar actuator system, showing the rigid body decoupled transfer in $z$-direction for several \emph{frozen positions} of the mover.}
    \label{fig:Motioncontrolloop}
\end{figure}

Many high-precision motion systems exhibit position dependent effects, see Figure \ref{fig:Motioncontrolloop}, which are introduced by relative sensing and actuation of the moving body.
Therefore, such systems are often represented in an LPV form, see \cite{5714737}, where position dependency is expressed by a so-called scheduling variable. Consider the equations of motion of a mechatronic system that is subject to relative actuation and relative sensing of the moving-body:
\begin{equation}
    M\Ddot{q}(t) + D\dot{q}(t) + Kq(t) = \Phi_a (p(t)) u(t),
    \label{GeneralizedDynamics}
\end{equation}
where $M$, $D$ and $K$ are the real symmetric mass, damping and stiffness matrices of dimension $n_q\times n_q$ and $\Phi_a(p(t)) \in \mathbb{R}^{n_q\times n_u}$ maps the control forces $u(t)$ to the appropriate masses based on the scheduling vector $p:\mathbb{R}\rightarrow \mathbb{P}\subseteq\mathbb{R}^{n_p}$. Alternatively, (\ref{GeneralizedDynamics}) is represented in LPV state-space form as:
\begin{equation}
P = 
 \left[ \begin{array}{cc|c}
    0 & I & 0\\
    -M^{-1}K & -M^{-1}D & M^{-1}\Phi_a(p(t))
    \\
    \hline 
    \Phi_s(p(t)) &0 & 0
 \end{array}\right],
 \label{PlantDynamics}
\end{equation}

\noindent where $\Phi_s(p(t)) \in \mathbb{R}^{n_y\times n_q}$ corresponds to a mapping of $q(t)$ to the output $y(t)$ based on the scheduling vector $p(t)$.
Note that the $B(p(t))$ and $C(p(t))$  matrices are position dependent due to relative actuation and sensing of the moving-body. If the scheduling vector $p(t)$ is constant, implying $p(t) = {\tt p} \in \mathbb{P}$ for all $t \in \mathbb{R}$, (\ref{PlantDynamics}) becomes an LTI system, which is often referred to as \emph{frozen dynamics} for a particular fixed position of the motion system and is denoted by $P_{\tt p}$. Typically, such a family of \emph{frozen dynamics} is used in \emph{sequential loop closing} (SLC) control design strategies by viewing the position dependent effects as a dynamic uncertainty. Consequently, during conventional SLC control design, closed-loop performance is traded off against robustness in order to ensure local stability of the closed-loop system.  Moreover, conventional SLC control design strategies are no longer sufficient to meet with the increasing performance requirements in the industry.
 Therefore, there is a necessity for a (practical) control design approach that is able to push performance beyond LTI control design solutions, such that satisfactory closed-loop performance is achieved.

\subsection{Problem statement}
The problem that is addressed in this paper is to construct a position dependent control law that considers all local aspects of the system by following a loop-shaping based approach, capable to handle the MIMO nature of the plant. The objective is to design a controller $K$, such that the following requirements are satisfied:
\begin{itemize}
  \item[(R1)] The closed-loop system is locally stabilized by $K$ for all $\tt p \in \mathbb{P}$.
  \item[(R2)] The control design approach utilizes non-parametric models, e.g. FRF measurements, thus circumventing the necessity of identifying an accurate dynamic model of the system that can capture complicated high-frequent flexible dynamics.
   \item[(R3)] The control design approach allows for modular design of the resulting feedback controller $K$ such that a desirable structure can be imposed on the controller.
\end{itemize}

The goal of this paper is the development of an extension of the conventional LTI SLC control design framework towards a position dependent approach, such that the position dependent nature of the system is exploited during feedback control design in order to maximize closed-loop performance.

\section{Conventional SLC control design framework}
\label{Section_LPVSLCframework}
This section gives a brief overview of the conventional SLC control design framework (e.g. see \cite{skogestad2007multivariable}), which is widely employed in industrial applications. 
The key concept of the SLC framework is that the multi-variable control design is decomposed into a number of equivalent SISO control designs, for which well-known control design tools are available, e.g. loop-shaping techniques. The equivalent plant for the $i$-th design step of the \emph{frozen} high-precision motion system dynamics, denoted by $g_{\tt p}^i$, is described by:
\begin{equation}
    g_{\tt p}^i = \mathcal{F}(P_{\tt p}^i,-K^i),
    \label{LFT}
\end{equation}

\noindent where $K^i$ = diag($k_j$) with $j = \lbrace 1,\hdots,n_u \rbrace$ and $j \neq i$, see \cite{oomen2020model}. The diagonal feedback controller corresponds to $K=\text{diag}(k_1,\hdots,k_{n_u})$.
Furthermore, $P_{\tt p}^i = W_i  P_{\tt p}W_i$, where $W_i$ corresponds to an identity matrix which has its $1$-st and $i$-th row interchanged.  Closed loop stability of the multi-variable rigid body decoupled system is assessed by considering the characteristic equation $\mathrm{det}(I+P_{\tt p}K) \ \forall  \tt p \in \mathbb{P}$, which is related to the equivalent plant as (see \cite{mayne1979sequential}): 
\begin{equation}
    \mathrm{det}(I+P_{\tt p}K) = \prod_{i=1}^{n_{u}}(1+g_{\tt p}^i k_i) \ \forall  \tt p \in \mathbb{P}
    \label{Stability1}
\end{equation}

 
\noindent
In terms of (\ref{Stability1}), stability of the multi-variable system is reduced to the assessment of $n_u$ SISO Nyquist criteria.

Generally, the controllers $k_i$ are designed in an iterative manner using the rigid body decoupled dynamics of the corresponding high-precision motion system, such that (local) stability is achieved for all \emph{frozen positions} $\tt p \in \mathbb{P}$, thus sacrificing performance for robustness during  control design due to the presence of position dependent flexible dynamics. 
\begin{figure}[b]
\vspace*{-6mm}
    \centering
    \includegraphics[trim={0.6cm 0cm 0.6cm 0cm},clip,width=0.9\linewidth]{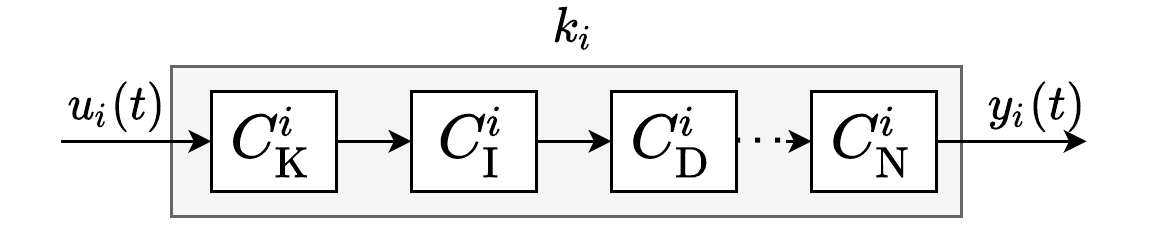}
    \caption{Typical structure of a SLC based rigid body feedback controller, where the controller is composed from a cascade interconnection of sub-components, such as a proportional gain, an integrator, a lead filter and notch filter.}
    \label{fig:cascade}
\end{figure}
Typically, a SLC based rigid body feedback controller is structured as a cascade interconnection of various LTI filters as illustrated by Figure \ref{fig:cascade}, where $C_\mathrm{K}^i$ corresponds to a proportional gain, which ensures that the open loop dynamics of the equivalent plant $g_{\tt p}^i k_i$ cross the 0 dB line at the desired rigid body target bandwidth. To suppress low-frequent disturbances, integral action is imposed on $k_i$ by means of the filter $C_\mathrm{I}^i$, which corresponds to:
\begin{equation}
    \begin{split}
        \dot{x}_\mathrm{I}(t) &= u_\mathrm{I}(t) \\
        y_\mathrm{I}(t) &= x_\mathrm{I}(t)
    \end{split}
    \label{Integrator}
\end{equation}

\noindent 
In order to stabilize the closed-loop system, lead filters are added to the SLC based controller. The time domain representation of a lead filter $C_\mathrm{D}^i$  is given by:
\begin{equation}
    \begin{split}
        \dot{x}_\mathrm{D}(t) &= -2\pi \alpha f_{\mathrm{bw}} x_\mathrm{D}(t) + 2\pi \alpha f_{\mathrm{bw}} u_\mathrm{D}(t) \\
        y_D(t) &= 1-\alpha^2x_\mathrm{D}(t) + \alpha^2u_\mathrm{D}(t)
    \end{split},
    \label{Differentiator}
\end{equation}

\noindent such that approximately 45 degrees phase margin is present at the rigid body target bandwidth $f_{\mathrm{bw}}>0$, while being subject to integral action $C_\mathrm{I}^i$. The parameter $\alpha>0$ is used to shift the cut-off frequencies of the differential action and the integral action of the lead filter, such that sufficient phase lead is present at the desired target bandwidth. Typically, $\alpha$ is chosen to be 3.
At last, a notch filter $C_\mathrm{N}^i$ is imposed on $k_i$ to suppress the effects of position dependent flexible dynamics. The time domain representation of a notch filter $C_\mathrm{N}^i$ is denoted by:
\begin{equation}
    \begin{split}
        \dot{x}_\mathrm{N}(t) &= \begin{bmatrix}-4\pi \beta_2 f_2 & -4f_2^2  \pi^2 \\ 1 & 0 \end{bmatrix}x_\mathrm{N}(t)+\begin{bmatrix}4f_2^2 \pi^2 \\0\end{bmatrix}u_\mathrm{N}(t) \\
        y_\mathrm{N}(t) &= \begin{bmatrix}\frac{\beta_1f_1-\beta_2f_2}{f_1^2 \pi} & 1-\frac{f_2^2}{f_1^2} \end{bmatrix}x_\mathrm{N}(t) + \frac{f_2^2}{f_1^2}u_\mathrm{N}(t)
    \end{split},
\end{equation}

\noindent 
where $\beta_1$, $\beta_2$, $f_1$ and $f_2$ correspond to the notch filter coefficients respectively.

\section{LPV SLC control design approach}
\label{LPVExten}
This section presents a novel extension of the conventional SLC control design framework towards a position dependent approach, which allows to push closed-loop performance of the position dependent system beyond LTI control strategies.

Similar to the SLC control design framework, the (local) equivalent plant for a particular frozen position of the motion system, denoted by $g_{\tt p}^i$, corresponds to a lower fractional transformation (LFT):
\begin{equation}
    g_{\tt p}^i = \mathcal{F}(P_{\tt p}^i,-K_{\tt p}^i),
    \label{LFT1}
\end{equation}

\noindent where $K_{\tt p}^i$ = diag(${k_{{\tt p}_j}}$), $j = \lbrace 1,\hdots,n \rbrace$, $j \neq i$. Moreover, the local diagonal feedback controller, which corresponds to the local plant dynamics $P_{\tt p}$, is defined as $K_{\tt p} = \text{diag}(k_{{\tt p}_1},\hdots,k_{{\tt p}_{n_u}})$, where local stability is assessed using the characteristic equation det($I+P_{\tt p}K_{\tt p}$) $\forall  \tt p \in \mathbb{P}$, which, in case of a position dependent controller, is related to the equivalent plant as:
\begin{equation}
    \mathrm{det}(I+P_{\tt p}K_{\tt p}) = \prod_{i=1}^{n_{\mathrm{u}}}(1+g_{\tt p}^i {k_{\tt p}}_i) \ \forall  \tt p \in \mathbb{P}
    \label{Stability}
\end{equation}

For the design of the position dependent controllers $k_{{\tt p}_i}$, the structured controller is partitioned into a low-frequent LTI controller, denoted by $\Gamma_i$, and an interconnection of LPV filters, denoted by  $\Psi_i$, as illustrated by Figure \ref{fig:ControlPartition}. Typically, $\Gamma_i$ consists of a proportional gain $C_\mathrm{k}^i$, an integrator $C_\mathrm{I}^i$, see (\ref{Integrator}), and lead filters $C_\mathrm{D}^i$, see (\ref{Differentiator}). The position dependent cascade filter interconnection $\Psi_i$ consists of LPV notch filters, which actively combat the effects of position dependent flexible dynamics in a position dependent manner, therefore allowing for increased rigid body feedback control bandwidth. The time-domain representation of a LPV notch filter corresponds to:
\begin{figure}[t]
    \centering
    \includegraphics[width=0.9\linewidth]{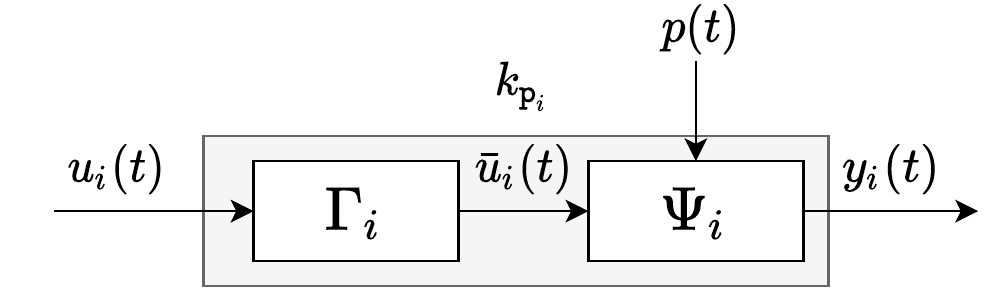}
    \caption{Schematic representation of the proposed LPV SLC control design concept, where $\Gamma_i$ denotes a cascade interconnection of LTI filters and $\Psi_i$ corresponds to a cascade interconnection of LPV filters.}
    \label{fig:ControlPartition}
\end{figure}
\begin{equation}
\left [
\begin{array}{c}
   \dot{x}_\mathrm{N}(t)\\ y_\mathrm{N}(t)
\end{array}
\right]
=
\left[ 
\begin{array}{c|c}
     \mathcal{A}(p(t)) & \mathcal{B}(p(t))  \\ 
     \hline 
     \mathcal{C}(p(t))  & \mathcal{D}(p(t)) 
\end{array}
\right]
\left[
\begin{array}{c}
     x_\mathrm{N}(t)\\  
     u_\mathrm{N}(t)
\end{array}
\right],
\label{notcher}
\end{equation}

\noindent with:
\begin{small}
\begin{equation}
    \begin{split}
        \mathcal{A}(p(t))  &= \begin{bmatrix}-4\pi \beta_2(p(t)) f_2(p(t)) & -4f_2(p(t))^2  \pi^2 \\ 1 & 0 \end{bmatrix}\\
        \mathcal{B}(p(t))  &= \begin{bmatrix}4f_2(p(t))^2 \pi^2 \\0\end{bmatrix}\\
        \mathcal{C}(p(t))  &= \begin{bmatrix}\frac{\beta_1(p(t))f_1(p(t))-\beta_2(p(t))f_2(p(t))}{f_1(p(t))^2 \pi} & 1-\frac{f_2(p(t))^2}{f_1(p(t))^2} \end{bmatrix} \\
        \mathcal{D}(p(t))  &= \frac{f_2(p(t))^2}{f_1(p(t))^2}
    \end{split}
\end{equation}
\end{small}
Similar to the SLC control design approach, a set of \emph{frozen dynamics}, denoted by $\{P_{\tt p}^l\}_{l=1}^{n_g}$, is used to construct a set of \emph{frozen controllers} $\{K_{\tt p}^l\}_{l=1}^{n_g}$ by utilizing loop shaping techniques, thus allowing for an interpolation based parameterization of the position dependent filter coefficients using the corresponding \emph{frozen scheduling vectors} $\tt p \in \mathbb{P}$ respectively, where $n_g$ denotes the number of \emph{frozen positions} that are considered for controller design. The parameterization of the position dependent filters can be accomplished using two approaches. First, a global parameterization of the filter coefficients can be considered with polynomial dependence on the parameters. Therefore, the polynomial dependence can be tuned using SLC control design approaches , thus taking all the local dynamics into account during the tuning of the position dependent filters. As an alternative, local parameterization strategies can be applied, for which each individual filter coefficient is interpolated based on a polynomial relationship which is dependent on the scheduling vector. In this paper, the latter strategy is investigated. 

Consider the vector $\Phi(p(t))$, which contains the position dependent notch filter coefficients of (\ref{notcher}):
\begin{equation}
\begin{small}
    \Phi(p(t)) = \begin{bmatrix}\beta_1(p(t)) & \beta_2(p(t))&f_1(p(t)) & f_2(p(t))\end{bmatrix}^\top
    \end{small}
\end{equation}

\noindent Using the set of local controller designs $\{K_{\tt p}^l\}_{l=1}^{n_g}$, the position dependent filter coefficients can be individually parameterized using a polynomial dependence on the scheduling vector as:
\begin{equation}
        \Phi_m(p(t)) = \sum_{v=0}^{(i-1)} \sum_{w=0}^{(j-1)} \theta_{vw}^{\Phi_m} q_x(t)^{v} q_y(t)^{w}, 
    \label{PolynomialPar}
\end{equation}

\noindent where $\Phi_m(p(t))$ denotes the $m^{\mathrm{th}}$ element of the vector $\Phi(p(t))$ and $q_x,q_y \subseteq \tt p$. The parameters $i$ and $j$ denote the order of the assumed dependency, governing the smoothness of the allowed variation. Additionally, from (\ref{PolynomialPar}), it is observed that each parameter $\Phi_m(p(t))$ is expressed by spatial coordinates (e.g. dependency on the $q_x(t),q_y(t)$ position) and weighting coefficients $\theta_{vw}^{\Phi_m}$, thus allowing for reformulation of the parameterization as:
\begin{equation}
     \Phi_m(p(t)) = \chi(p(t)) \begin{bmatrix}
        \theta_{11}^{\Phi_m} & \hdots & \theta_{ij}^{\Phi_m}
        \end{bmatrix}^\top,
        \label{TruncatedParameterization}
\end{equation}

\noindent where the vector $\chi(p(t)) \in \mathbb{R}^{i \cdot j}$ corresponds to:
\begin{equation}
    \chi(p(t)) = \begin{bmatrix}
1&\hdots&q_x(t)^{i-1}
\end{bmatrix} \otimes \begin{bmatrix}
1&\hdots&q_y(t)^{j-1}
\end{bmatrix} 
\end{equation}

\noindent
In order to obtain the weighting coefficient vectors, the polynomial parameterization, expressed by (\ref{TruncatedParameterization}), is reformulated into a least-squares regression problem using the designed local filter coefficients of $\{K_{\tt p}^l\}_{l=1}^{n_g}$ and their corresponding \emph{frozen positions} respectively. Moreover, the least-squares formulation for the approximation of $\Phi_m(p(t))$ is given by:
\begin{equation}
\underbrace{
    \begin{bmatrix}
    {\Phi_m}^1 \\ \vdots \\{\Phi_m}^{n_g}
    \end{bmatrix}}_{Y_{\Phi_m}}
    =
    \underbrace{
    \begin{bmatrix}
    \chi({\tt p}_1) \\ \vdots \\ \chi({\tt p}_{n_g})
    \end{bmatrix}}_{A}
    \underbrace{
    \begin{bmatrix}
    \theta_{11}^{\Phi_m} \\ \vdots \\ \theta_{ij}^{\Phi_m}
    \end{bmatrix} }_{\Theta_{\Phi_m}},
     \label{LSR}
\end{equation}

\noindent where the coefficient vectors $\Theta_{\Phi_m}$ are obtained by minimizing $||A\Theta_{\Phi_m}-Y_{\Phi_m}||^2$. Therefore, the parameterization of the position dependent filter coefficients is based on minimizing the $l_2$ loss between the local coefficient tuning of $\{K_{\tt p}^l\}_{l=1}^{n_g}$ and the fit of the polynomial parameterization.

\begin{figure}[h]
    \centering
    \includegraphics[trim={0.6cm 0cm 0.6cm 0cm},clip,width=\linewidth]{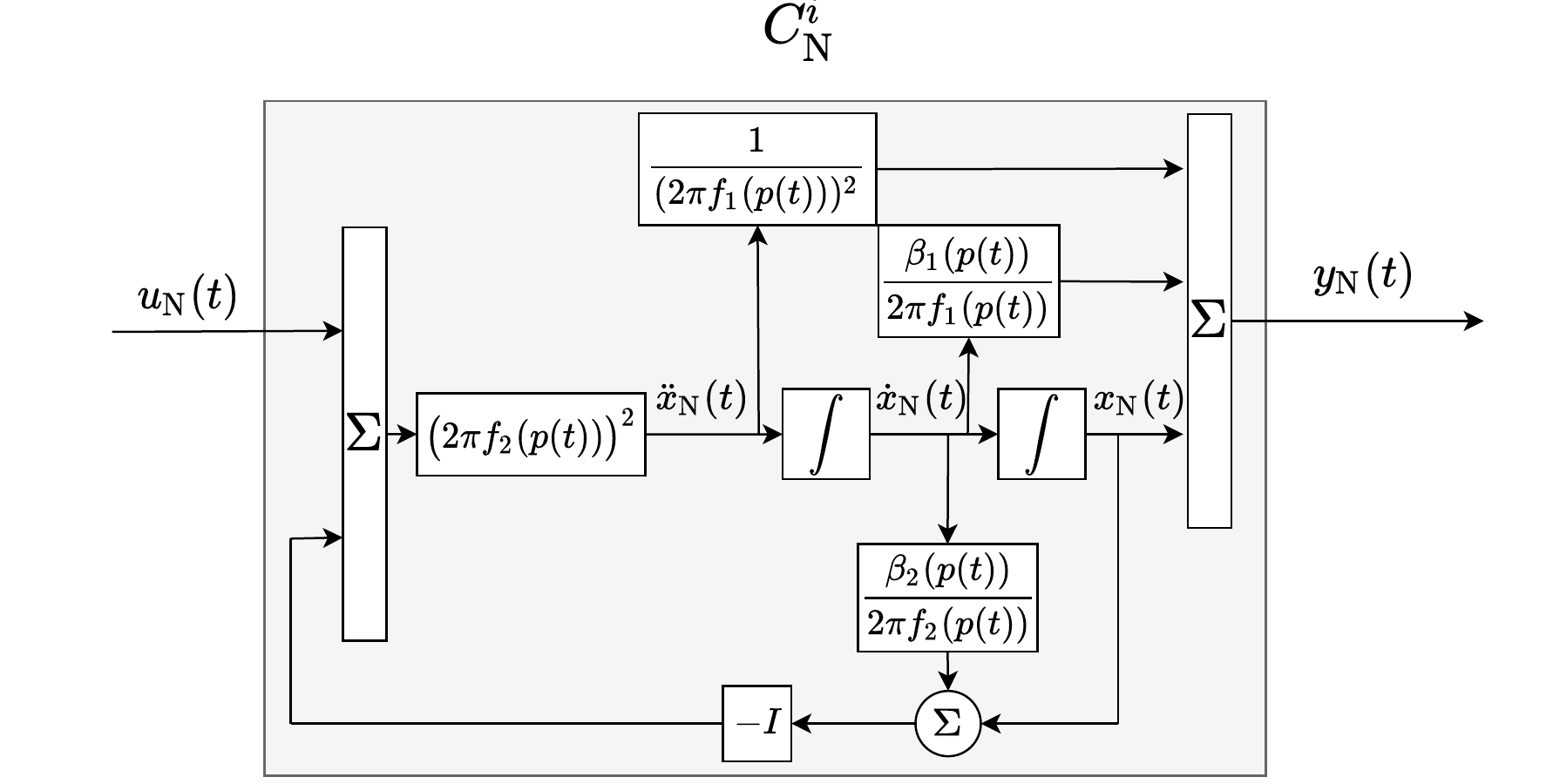}
    \caption{Implementation concept of a position dependent notch filter using an integrator chain.}
    \label{fig:PositionNotch}
\end{figure}

Real-time implementation of the position dependent controller blocks is achieved by using the controller structure given by (\ref{notcher}), which allows for implementation of the position dependent notch filter using an integrator chain
with position dependent coefficients as illustrated in Figure \ref{fig:PositionNotch}.

In total, a novel control design approach has been proposed, which allows for position dependent sequential loop closing based control design. In this way, position dependent phenomena in high-precision motion systems can be explicitly handled, allowing for increased closed-loop performance. Additionally, the structure of the position dependent filters allows for efficient implementation of the filters using an integrator chain, where the total feedback controller $k_i$ is obtained by interconnecting the LTI filters $\Gamma_i$ with the LPV filters $\Psi_i$. 
\section{Simulation study}
\label{Section_Academicexample}

In this Section, a simulation study of the LPV SLC control design approach is presented. First Subsection \ref{subsection_SystemDescription} presents a brief description of a moving-magnet planar actuator system. Next, simulation results are presented in Subsection \ref{subsection_simulation}, where the proposed LPV SLC control design approach is compared to conventional SLC control design strategies on a state-of-the-art moving-magnet planar actuator model, which exhibits position dependency in both actuation and sensing.
\subsection{System description}
\label{subsection_SystemDescription}
\begin{figure}[t]
    \centering
    \includegraphics[width=\linewidth]{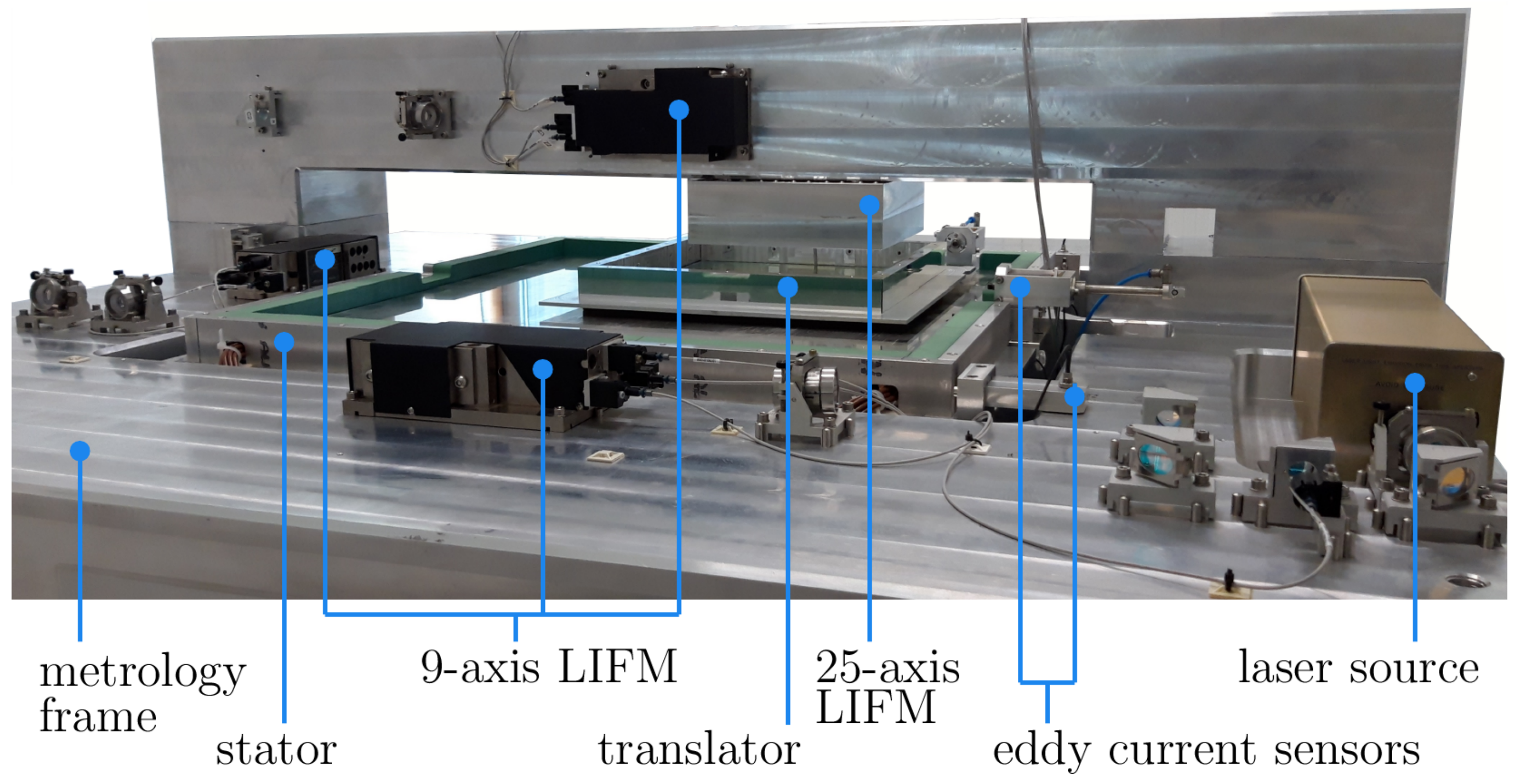}
    \caption{Photograph of a moving-magnet planar actuator system prototype.}
    \label{fig:MMPA}
\end{figure}

Magnetically levitated planar actuators are high-precision motion systems which are used in lithographic machines for nanometer accurate positioning of silicon wafers under projection optics, see \cite{Butler}. A moving-magnet planar actuator system, which is illustrated by Figure \ref{fig:MMPA}, is comprised of three main components: the stator base, the translator and the metrology frame.  The stator base is a double layer coil array, consisting of 160 coils of which 40 coils are simultaneously activated at every time instant using 40 power converters, depending on the relative position of the translator (see \cite{Lierop-phd}).  Proper actuation of the coils offers the means of both levitation and propulsion of the magnet plate in 6 Degrees of Freedom (DoF). The translator, comprised of 281 permanent magnets structured in a Hallbach array, is constructed to be a lightweight magnetic plate, thus enabling high accelerations. However, due to its low mass, low-frequent flexible dynamics are introduced (first flexible mode at $\approx$ 226.5 Hz), which severely limit position tracking performance capabilities. The metrology frame, which rests on air mounts to suppress the effects of floor disturbances, is used as a global reference frame in order to assess positioning accuracy of the mover. On the metrology frame, 9 laser interferometers (LIFMs) are mounted to measure the relative displacement of the translator with respect to the metrology frame. The relative actuation and sensing of the moving-body introduces the presence of position dependent flexible dynamics, which severely limit the achievable rigid body feedback control bandwidth, see Figure \ref{fig:Motioncontrolloop}. A detailed description of the moving-magnet planar actuator prototype is given in \cite{8785302}.

\subsection{Simulation results}
\label{subsection_simulation}

\begin{figure}[h]
    \centering
    \includegraphics[trim={2cm 0cm 2.1cm 0cm},clip,width=\linewidth]{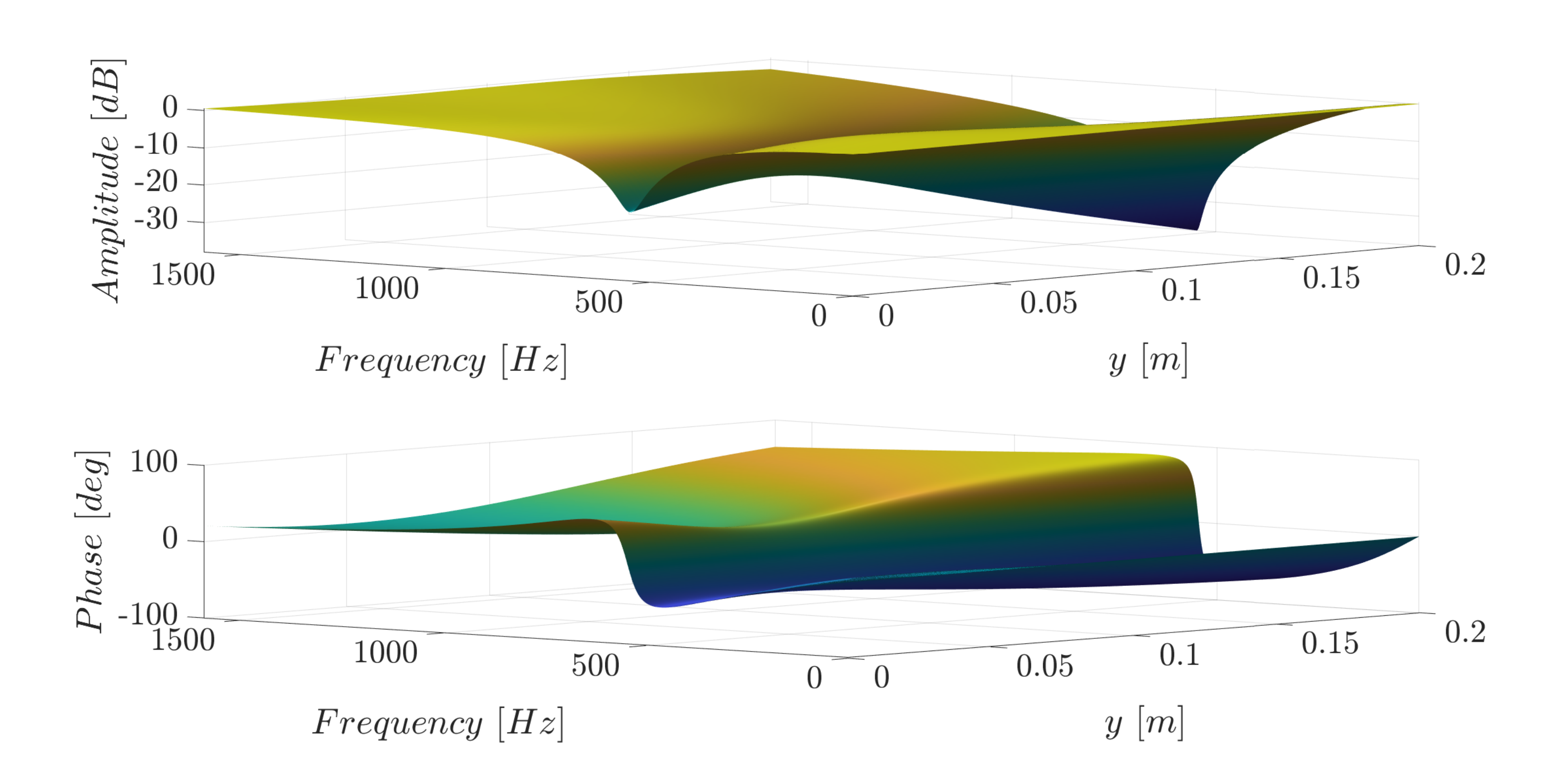}
    \caption{Visualization of a designed position dependent notch filter for the positions $q_x$ = 0.1$m$ and $q_y \in [0m\quad 0.2m]$.}
    \label{fig:positiondependentnotch}
\end{figure}

For validation of the proposed control design approach, both a conventional SLC feedback controller and a LPV SLC controller are designed for the $z$-axis of the moving-magnet planar actuator model, which is severely limited in achieving sufficient rigid body feedback control bandwidth due to the presence of position dependent flexible dynamics as illustrated in Figure \ref{fig:Motioncontrolloop}. In order to compare the performance of both controllers with respect to each other, a performance criteria is introduced by means of a 6 dB upper-bound on the sensitivity function of the closed-loop. Additionally, both controllers are constructed from a proportional gain, an integrator, lead filters and (position dependent) notch filters. Using the aforementioned filters, the conventional SLC feedback controller achieves a rigid body feedback control bandwidth of approximately 95 Hz. The LPV SLC feedback controller achieves a feedback control bandwidth of approximately 175 Hz due to the additional degrees of freedom during controller design. The introduction of position dependent notch filters to the controller design has several advantageous properties. First, the effects of position dependent flexible dynamics can be explicitly handled in a position dependent manner, thus allowing to push performance beyond robust LTI SLC control design. Secondly, the position dependent notch filters allow for relaxation of the feedback control design constraints, since phase lead can be introduced in a position dependent manner by means of skewing the notch filter (shifting the notch filter frequency $f_2$, see (\ref{notcher}), to a higher frequency compared to $f_1$), thus allowing for increased feedback control bandwidth, while still satisfying the 6 dB upper-bound of the sensitivity function respectively. The constructed position dependent notch filter is illustrated in Figure \ref{fig:positiondependentnotch} for the positions $q_x$ = 0.1 $m$, $q_y\in [0m\quad 0.2m]$.

\begin{figure}[h]
    \centering
    \includegraphics[trim={3cm 1.2cm 3cm 0cm},clip,width=\linewidth]{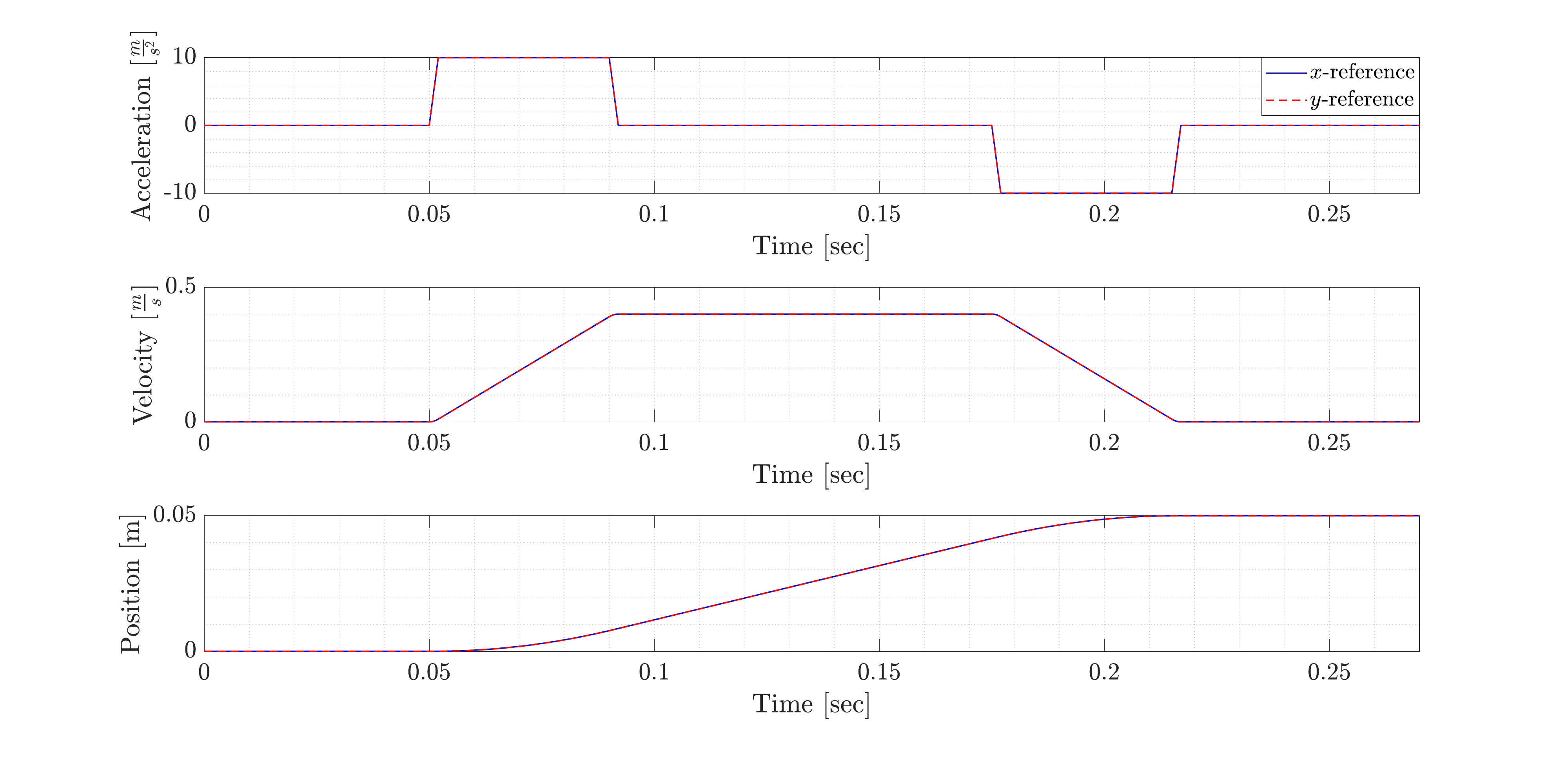}
    \caption{Motion profiles that are considered for simulation of the conventional SLC controller and the LPV SLC controller.}
    \label{fig:ReferenceTraj}
\end{figure}

From Figure \ref{fig:positiondependentnotch} it is observed that the constructed position dependent notch filter allows for introduction of phase lead for critical positions of the system due to the additional design freedom of the position dependent filters, thus allowing for increased closed-loop performance compared to conventional SLC control design approaches. Additionally, the designed notch filter is suppressing the effects of the resonance dynamics. In order to evaluate the performance improvement of the proposed LPV SLC control design approach, both the conventional SLC controller and the LPV SLC controller are implemented in simulation together with a mass feedforward. For simulation purposes, a 4$^{\mathrm{th}}$ order reference trajectory is generated based on the methodology proposed by \cite{208Lambrechts}. Figure \ref{fig:ReferenceTraj} illustrates the motion profiles that are used for simulation.

\newpage 
\begin{figure}[h]
    \centering
    \includegraphics[trim={3.2cm 0cm 3.2cm 0cm},clip,width=\linewidth]{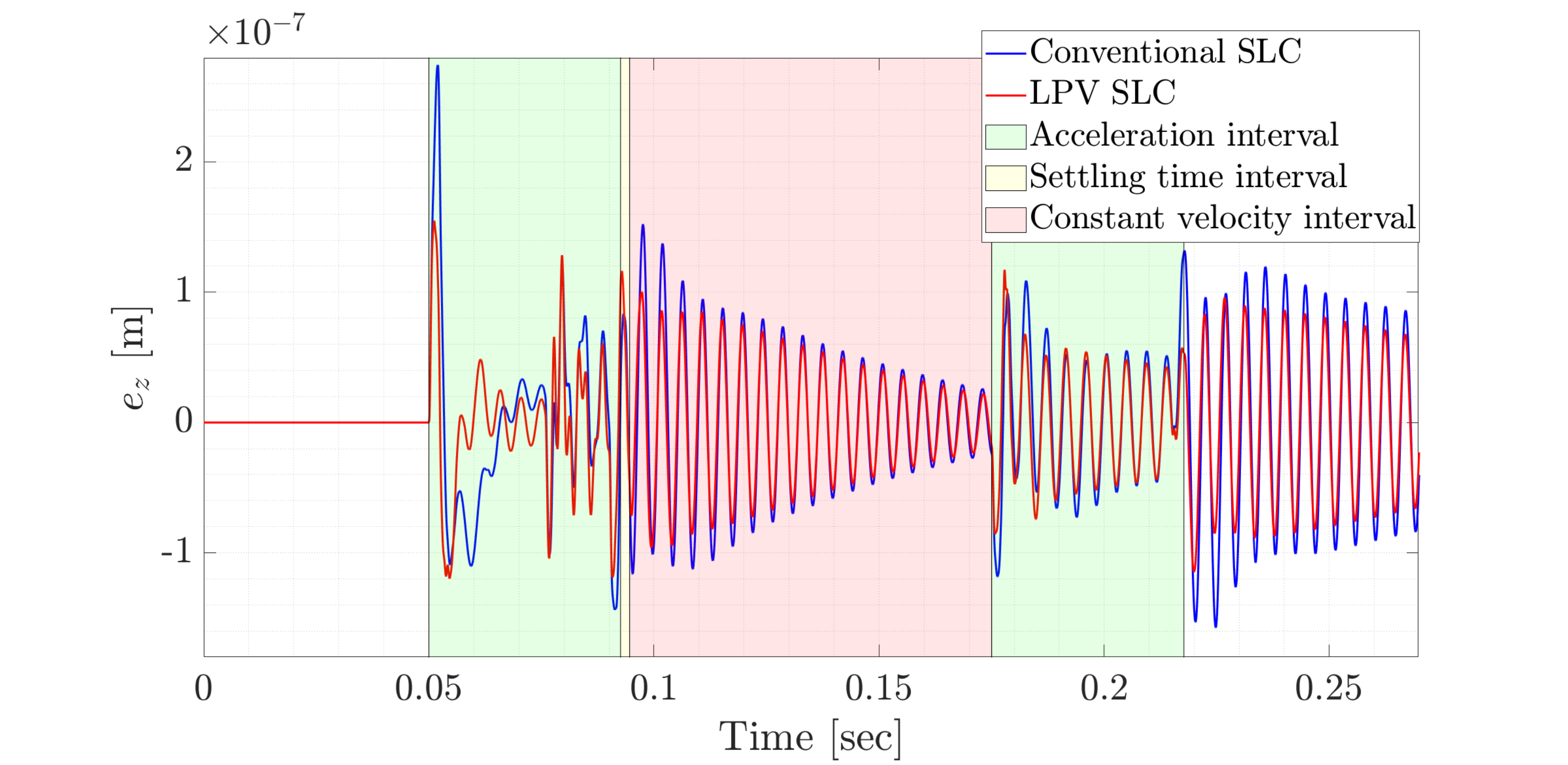}
    \caption{Position tracking error in $z$-direction, where the green area denotes the (de-)acceleration interval, the yellow area denotes the settling time interval and the red area denotes the constant velocity interval.}
    \label{fig:ztrackignerror}
\end{figure}

The resulting tracking error is illustrated in Figure \ref{fig:ztrackignerror}, where the position
tracking error using the conventional SLC controller is depicted in blue and the position tracking error using the LPV SLC controller is depicted in red. Furthermore, the green area denotes the acceleration and de-acceleration interval, the yellow area denotes the settling time interval and the red area denotes the constant velocity interval.

In industry, accuracy of the lithographic exposure process, which takes place during the constant velocity interval, is assessed by two values: the Moving Average ($\mathrm{MA}(t)$) and the Moving Standard Deviation ($\mathrm{MSD}(t)$), see \cite{Butler}. These values are calculated a posteriori from the measured position tracking error using:
\begin{equation}
    \begin{split}
        \mathrm{MA}(t) &= \frac{1}{T}\int_{t-\frac{T}{2}}^{t+\frac{T}{2}} e(\tau)d\tau\\
        \mathrm{MSD}(t) &= \sqrt{\frac{1}{T}\int_{t-\frac{T}{2}}^{t+\frac{T}{2}} (e(\tau)-\mathrm{MA}(t))^2d\tau}
    \end{split},
\end{equation}


\noindent 
where $T$ corresponds to the exposure window. The mean values of the $\mathrm{MA(t)}$ and $\mathrm{MSD(t)}$ filtered error signals during the constant velocity interval are listed in Table \ref{Tabelja} together with the relative reduction of the performance values.

\begin{table}[h]
\caption{Mean values of $\mathrm{MA(t)}$ and $\mathrm{MSD(t)}$ for both controllers during the constant velocity interval}
\label{Tabelja}
\vspace*{-2mm}
\begin{center}
\begin{tabular}{|c|c|c|c|}
\hline
                                      & \textbf{Conventional SLC} & \textbf{LPV SLC}    & \textbf{Relative reduction} \\ \hline
\textbf{MA} {[}m{]}  & 5.3139e-10       & 4.4360e-11 & 91.65 \%           \\ \hline
\textbf{MSD} {[}m{]} & 4.8913e-08       & 4.0785e-08 & 16.62 \%           \\ \hline
\end{tabular}
\end{center}
\end{table}

\vspace*{-2mm}
\noindent Based on Table \ref{Tabelja}, it is clear that the LPV SLC design based controller has a relative reduction of the mean $\mathrm{MA(t)}$ error of 91.65 \% and a relative reduction of the mean $\mathrm{MSD}(t)$ error of 16.62 \%, thus highlighting the effectiveness of the proposed control design approach. From Figure \ref{fig:ztrackignerror} it is difficult to observe the improvement in $\mathrm{MA}(t)$ error as the low-frequency behavior is obscured by the resonance. Moreover, an important observation is that the LPV SLC controller is better equipped to handle the position dependent resonance, which leads to increased feedback control bandwidth. Nonetheless, the resonant behavior ($\mathrm{MSD}$ error) is hardly reduced as it is excited by the mass feedforward.

\section{Conclusions}
\label{Section_Conclusion}
This paper presents a novel extension of the conventional SLC control design framework towards a linear-parameter-varying approach, which allows to circumvent limitations that are introduced by position dependent effects of high-precision motion systems. Moreover, the proposed framework introduces additional degrees of freedom to the controller design by means of exploiting position dependent filter design, which allows for increased closed-loop performance of high-precision motion systems. Simulation results based on a state-of-the-art moving-magnet planar actuator model highlight the effectiveness of the proposed control design approach. 

\addtolength{\textheight}{-12cm}   




\bibliographystyle{ieeetr}       
\bibliography{MyBib}

\begin{thebibliography}{10}

\bibitem{Butler}
H.~Butler, ``Position control in lithographic equipment [applications of
  control],'' {\em IEEE Control Systems Magazine}, vol.~31, no.~5, pp.~28--47,
  2011.

\bibitem{411117}
N.~Tamer and M.~Dahleh, ``Feedback control of piezoelectric tube scanners,'' in
  {\em Proceedings of 1994 33rd IEEE Conference on Decision and Control},
  vol.~2, pp.~1826--1831 vol.2, 1994.

\bibitem{HEERTJES20161}
M.~Heertjes, ``Data-based motion control of wafer scanners,'' {\em
  IFAC-PapersOnLine}, vol.~49, no.~13, pp.~1--12, 2016.
\newblock 12th IFAC Workshop on Adaptation and Learning in Control and Signal
  Processing ALCOSP 2016.

\bibitem{6225187}
X.~Ye, Y.~Zhang, and Y.~Sun, ``Robotic pick-place of nanowires for
  electromechanical characterization,'' in {\em 2012 IEEE International
  Conference on Robotics and Automation}, pp.~2755--2760, 2012.

\bibitem{Oomen}
T.~Oomen, ``Advanced motion control for precision mechatronics: control,
  identification, and learning of complex systems,'' {\em IEEJ Journal of
  Industry Applications}, vol.~7, pp.~127--140, Jan. 2018.

\bibitem{Steinbuch2013}
M.~Steinbuch, ``Design and control of high tech systems,'' in {\em 2013 IEEE
  International Conference on Mechatronics (ICM)}, pp.~13--17, 2013.

\bibitem{190Steinbuch}
M.~Steinbuch, R.~Merry, M.~Boerlage, M.~Ronde, and M.~{Molengraft, van de},
  {\em Advanced Motion Control Design}, pp.~27--1/25.
\newblock CRC Press, 2010.

\bibitem{HOVD19941601}
M.~Hovd and S.~Skogestad, ``Sequential design of decentralized controllers,''
  {\em Automatica}, vol.~30, no.~10, pp.~1601--1607, 1994.

\bibitem{VibrationModes}
P.~Hughes, ``Space structure vibration modes: How many exist? which ones are
  important?,'' {\em IEEE Control Systems Magazine}, vol.~7, no.~1, pp.~22--28,
  1987.

\bibitem{STEINBUCH1998278}
M.~Steinbuch and M.~Norg, ``Advanced motion control: An industrial
  perspective,'' {\em European Journal of Control}, vol.~4, no.~4,
  pp.~278--293, 1998.

\bibitem{skogestad2007multivariable}
S.~Skogestad and I.~Postlethwaite, {\em Multivariable feedback control:
  analysis and design}, vol.~2.
\newblock Citeseer, 2007.

\bibitem{5714737}
R.~Toth, H.~S. Abbas, and H.~Werner, ``On the state-space realization of lpv
  input-output models: Practical approaches,'' {\em IEEE Transactions on
  Control Systems Technology}, vol.~20, no.~1, pp.~139--153, 2012.

\bibitem{oomen2020model}
T.~Oomen and M.~Steinbuch, ``Model-based control for high-tech mechatronic
  systems,'' in {\em Mechatronics and Robotics}, pp.~51--80, CRC Press, 2020.

\bibitem{mayne1979sequential}
D.~Q. Mayne, ``Sequential design of linear multivariable systems,'' in {\em
  Proceedings of the Institution of Electrical Engineers}, vol.~126,
  pp.~568--572, IET, 1979.

\bibitem{Lierop-phd}
C.~M.~M. van Lierop, {\em Magnetically levitated planar actuator with moving
  magnets: Electromechanical analysis and design II}.
\newblock PhD thesis, Eindhoven University of Technology, 2007.

\bibitem{8785302}
C.~Custers, I.~Proimadis, J.~W. Jansen, H.~Butler, R.~Tóth, E.~Lomonova, and
  P.~van~den Hof, ``Active compensation of the deformation of a magnetically
  levitated mover of a planar motor,'' in {\em 2019 IEEE International Electric
  Machines Drives Conference (IEMDC)}, pp.~854--861, 2019.

\bibitem{208Lambrechts}
P.~Lambrechts, M.~Boerlage, and M.~Steinbuch, ``Trajectory planning and
  feedforward design for electromechanical motion systems,'' {\em Control
  Engineering Practice}, vol.~13, no.~2, pp.~145--157, 2005.

\end{thebibliography}

\end{document}